\pdfoutput=1
\documentclass[aps,pre,superscriptaddress,amsmath,amssymb,reprint,nofootinbib,twoside]{revtex4-2}
\usepackage{graphicx,bm,bbding}

\newcommand{\bigstaropen}{\mbox{\protect\raisebox{-0.55ex}{\FiveStarOpen}}}
\newcommand{\blackbigcirc}{\mbox{\protect\raisebox{-0.55ex}{{\huge$\bullet$}}}}
\newcommand{\phZ}{\phantom{0}}
\newcommand{\phM}{\phantom{$-$}}

\begin{document}

\title{Correspondence between classical and quantum resonances}
\author{F. J. Arranz}
\email{fj.arranz@upm.es}
\author{R. M. Benito}
\email{rosamaria.benito@upm.es }
\affiliation{Grupo de Sistemas Complejos, Universidad Polit\'ecnica de Madrid, av.\ Puerta de Hierro 2--4, 28040 Madrid, Spain}
\author{F. Borondo}
\email{f.borondo@uam.es}
\affiliation{Instituto de Ciencias Matem\'aticas (ICMAT), Cantoblanco, 28049 Madrid, Spain}
\affiliation{Departamento de Qu\'\i mica, Universidad Aut\'onoma de Madrid, Cantoblanco, 28049 Madrid, Spain}
\date{\today}

\begin{abstract}
Bifurcations take place in molecular Hamiltonian nonlinear systems as the excitation energy increases, this leading to the appearance of different classical resonances. In this paper, we study the quantum manifestations of these classical resonances in the isomerizing system CN-Li$\leftrightarrows$Li-CN. By using a correlation diagram of eigenenergies versus Planck constant, we show the existence of different series of avoided crossings, leading to the corresponding series of quantum resonances, which represent the quantum manifestations of the classical resonances. Moreover, the extrapolation of these series to $\hbar=0$ unveils the correspondence between the bifurcation energy of classical resonances and the energy of the series of quantum resonances in the semiclassical limit $\hbar\to0$. Additionally, in order to obtain analytical expressions for our results, a semiclassical theory is developed.
\end{abstract}

\maketitle

\section{\label{sec:intro}Introduction}

Since Bohr's correspondence principle was first introduced~\cite{Bohr.correspondence.principle}, the issue of the correspondence between classical and quantum mechanics has led to countless publications, still remaining as a current research topic of interest. In the case of integrable systems, the correspondence between both mechanics was established by Einstein~\cite{Einstein.quantization}, Brillouin~\cite{Brillouin.quantization}, and Keller~\cite{Keller.quantization} in the EBK quantization~\cite{Gutzwiller.caos}, which is based on the action quantization of phase space tori.

Accordingly, this topic continues to generate publications in different fields of physics, particularly in the case of non-integrable chaotic systems where the tori structure of phase space is total or partially destroyed~\cite{Gutzwiller.caos,Reichl.caos,Lichtenberg.Lieberman.caos}. Abundant literature regarding quantum-classical correspondence has recently been published in connection with terahertz generation in a gas-phase medium~\cite{terahertz.emission}, Rydberg-atom spectroscopy in amplitude-modulated optical lattices~\cite{Rydberg.spectroscopy}, cavity optomechanical systems involving an optical parametric amplifier~\cite{cavity.optomechanical}, out-of-time order correlators in quantum chaos~\cite{OTOC.quantum.chaos}, transport properties in Heisenberg spin chains~\cite{spin.transport}, thermodynamic entropy~\cite{thermodynamic.entropy}, Gross-Pitaevski model of Bose-Einstein condensate~\cite{BE.condensate}, quantum cromodynamics~\cite{quantum.cromodynamics}, and quantum fields~\cite{quantum.fields}, among other topics.

Classical resonances, in the sense of the Poincar\'e-Birkhoff theorem~\cite{Gutzwiller.caos,Reichl.caos,Lichtenberg.Lieberman.caos} for Hamiltonian nonlinear systems, are associated to periodic orbits. Moreover, periodic orbits have been shown to play a very important role in the quantum-classical correspondence. In the case that the system is not ergodic, this role is mediated by the remnants of invariant tori, that can be quantized using the EBK prescription~\cite{Zembekov.LiCN.semiclassical}. However, for fully chaotic systems, Gutzwiller~\cite{Gutzwiller.trace.formula} obtained, by means of his trace formula~\cite{Gutzwiller.caos,Reichl.caos}, a relationship where quantum eigenenergies are expressed in terms of a summation over all periodic orbits in the system. Currently, Gutzwiller's trace formula remains as the only general tool available to establish a quantitative connection between classical and quantum mechanics for chaotic systems. This trace formula also indicates the significance of periodic orbits in the quantum-classical correspondence phenomenon known as {\em scarring}, which was first studied by Heller~\cite{Heller.scars}. He coined the term {\em scar} to refer to the accumulation of probability density along isolated unstable periodic orbits. That is, periodic orbits are not only significant for eigenenergies (trace formula), but also for eigenfunctions (scarring). The idea that periodic orbits are the underlying hidden skeleton supporting quantum eigenstates is strongly endorsed by the results obtained using basis sets of scar functions~\cite{Revuelta.scar.basis}, or more generally, basis sets of functions localized over (stable and unstable) periodic orbits~\cite{Revuelta.PO.localized.basis}, where it has been shown that each eigenfunction is mostly expressed in terms of a few periodic orbits (i.e., a few basis functions localized over the corresponding periodic orbits).

On the other hand, the concept of quantum resonances, first proposed by Fermi in order to give a theoretical explanation for, at the time called, anomalous effect in the CO$_2$ Raman spectrum~\cite{Fermi.resonance}, went beyond the explanation of this effect, and the theory developed by this author has been shown to be of general application in quantum mechanics, in particular, it is intimately related to the notion of avoided crossing between eigenstates, when a parameter in the corresponding Hamiltonian is tuned. If this happens, the involved states are mixed by means of an orthogonal, or more generally unitary, transformation. When two eigenstates belong to the same irreducible representation of the symmetry group of the system is applicable the non-crossing rule of von Neumann and Wigner~\cite{vonNeumann.Wigner.non-crossing}, leading to the corresponding avoided crossing. Conversely, if the eigenstates have different symmetry, then crossing is allowed. In a typical avoided crossing of a Hamiltonian nonlinear system, an exchange of character between both involved eigenstates, usually defined in terms of approximate quantum numbers when those can be assigned, takes place. This change of character can also be quantitatively described by calculating the corresponding coupling matrix element, or alternatively the associated mixing angle which results from the integration of that coupling \cite{Arranz.LiCN.hbar.correlation1}.

Moreover, Roberts and Jaff\'e~\cite{Jaffe.quantum.classical.resonances} developed an interesting theory in this field by studying the correspondence between classical resonances and Fermi resonances, obtaining that the classical dynamics corresponding to a quantum state involved in a Fermi resonance needs not exhibit resonant behavior, and furthermore, the fact that the classical dynamics corresponding to a quantum state is resonant does not necessarily implies that the quantum state is involved in a Fermi resonance. However, these authors found a relationship between classical and quantum resonances, namely, if two quantum states exhibit an $n$:$m$ Fermi resonance, then the classical dynamics associated with the matrix elements connecting these two states will exhibit an $n$:$m$ classical resonance.

In our case, the consideration of variations of the value of the Planck constant leads to a correlation diagram of eigenenergies versus that parameter, and this will be the basis of this paper. Variation of the Planck constant can be considered as an awkward mathematical artifact, but notice that, regarding the calculation of Hamiltonian eigenvalues, this variation is mathematically equivalent to an (inverse) variation of the system masses. In the case of a molecular system, a similar effect can be attained by considering different isotopic compositions. Indeed, the procedure introduced in our correlation diagram can tune EBK quantized actions to the action of the classical resonance, allowing to observe avoided crossings which could otherwise be hidden at $\hbar=1$ a.u. due to the characteristics of the system. This method has been proven in the past very useful for the analysis of the eigenstates of different molecular systems~\cite{Arranz.LiCN.hbar.correlation1,Arranz.LiCN.hbar.correlation2,Arranz.LiCN.HCN.HO2.hbar.correlation,Parraga.KCN.hbar.correlation}. The crucial point is that our correlation diagram has been able to identify, for a number of floppy molecular systems including HCN, LiCN, KCN and HO$_2$, a frontier separating two very different regions in it. The region above the frontier consists of a myriad of overlapping and broad avoided crossing, originating a large mix of individual states that give rise to ergodic eigenstates, according to the nodal pattern criterion of Stratt, Handy and Miller~\cite{Stratt.nodal.pattern.criterion}. This was taken as an indication of quantum stochasticity by Marcus~\cite{Noid.overlapping.ACs}, thus constituting this region the proper niche or ecosystem for Heller's scars. On the contrary, in the region below the frontier, the avoided crossings are isolated and sharp, such that every curve in the correlation diagram can be more easily understood in terms of EBK quantization of tori remnants. In terms of Marcus' classification this is a region of order or regularity in the quantum sense. More interestingly, it was found in Ref.~\cite{Arranz.LiCN.hbar.correlation2} that the frontier separating the two regions is located in the area of the correlation diagram where localized wavefunctions due to scarring first start to appear.

In this paper, we present a detailed study of the regular region, i.e., below the frontier of scars, of the eigenenergies correlation diagram using the Plank constant as parameter for the LiCN molecular system. Different series of sharp and isolated avoided crossings, corresponding to the main quantum resonances, will be obtained. Also, the correspondence with the classical resonances, i.e., periodic orbits, is analyzed. Finally, semiclassical expressions for the series of quantum resonances are given, which permit an analytical extrapolation to the $\hbar\to0$ limit. In this way, we are in the position to establish a quantitative correspondence between the classical bifurcation energies and the energy of the quantum resonances series in the semiclassical limit, which constitutes the main result of this work.

The organization of the paper is as follows. Section~\ref{sec:description.calculations} is devoted to the description of the Hamiltonian model used to represent the LiCN molecular system (Sec.~\ref{sec:system}), as well as the exposition of the classical (Sec.~\ref{sec:class_calculations}) and quantum (Sec.~\ref{sec:quant_calculations}) calculations performed on this model. Section~\ref{sec:results.discussion} is devoted to the joint presentation and discussion of the results obtained. In Sec.~\ref{sec:class_resonances} the main classical resonances that arise as energy increase are shown and the corresponding bifurcation energies are listed. In Sec.~\ref{sec:quant_resonances} the correlation diagram of eigenenergies versus Planck constant is depicted and explained, including the main quantum resonances corresponding to different series of avoided crossings. Also, the energies of the resonances in the semiclassical limit, which are estimated by linear extrapolation of the series, are listed and put in correspondence with the classical bifurcation energies. In Sec.~\ref{sec:semiclassical} a semiclassical theory is developed by considering different power series expansions in $\hbar$, leading to analytical expressions for the resonance energies in the semiclassical limit. Lastly, the paper is summarized and the conclusions reached are presented in Sec.~\ref{sec:conclusions}.

\section{\label{sec:description.calculations}System description and calculations}

\subsection{\label{sec:system}Hamiltonian model}

The system studied in this work corresponds to the purely vibrational dynamics of the lithium isocyanide molecule CNLi, taking the most abundant isotopic composition $^{12}$C, $^{14}$N, and $^{7}$Li. For simplicity, the rotational motion will be not considered. Since the C-N bond is much stronger than the interactions with Li atom, an adiabatic decoupling of the corresponding degree of freedom is adequate, such that the C-N bond length can be fixed at its equilibrium value, i.e., we study the relative motion of Li atom and CN group.

This system can be modeled in Jacobi coordinates by means of the Hamiltonian function
\begin{equation}
\label{eq:Hamiltonian}
H = \frac{P_R^2}{2\mu_1} + \frac{P_\theta^2}{2}\left( \frac{1}{\mu_1R^2} + \frac{1}{\mu_2r_\text{eq}^2} \right) + V(R,\theta),
\end{equation}
where $\mu_1 = m_\text{Li}(m_\text{C}+m_\text{N})/(m_\text{Li}+m_\text{C}+m_\text{N})$ and $\mu_2 = m_\text{C}m_\text{N}/(m_\text{C}+m_\text{N})$ are reduced masses ($m_\text{Li}$, $m_\text{C}$, and $m_\text{N}$ being the corresponding atomic masses), $r_\text{eq}=2.19$ a.u. is the fixed N-C equilibrium length, $R$ is the length between the CN group	center of mass and the Li atom, and $\theta$ is the angle formed by the corresponding $r_\text{eq}$ and $R$ directions (i.e., N$\rightarrow$C and $\genfrac{}{}{0pt}{}{\text{C}}{\text{N}}$$\rightarrow$Li, respectively). Thus, $\theta=0$ corresponds to the linear configuration Li-CN, and $\theta=\pi$ rad to the linear configuration CN-Li. Last, $P_R$ and $P_\theta$ are the corresponding conjugate momenta, and $V(R,\theta)$ is the potential energy function describing the vibrational interactions.

%%%%%%%%%%%%%%%%%
\begin{figure}[t]
\includegraphics{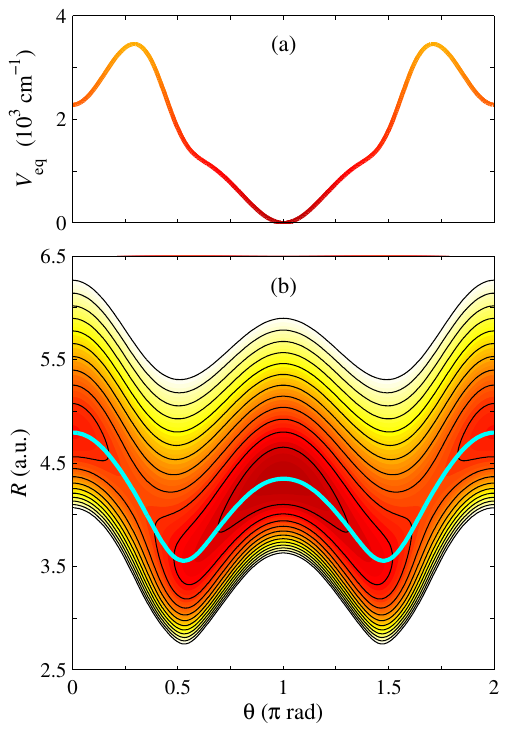}
\caption{\label{fig:pes_mep}%
(a) Energy profile along the minimum energy path connecting minima and saddles of the potential energy surface.
(b) Potential energy surface represented as contour plots spaced 1000 cm$^{-1}$. The minimum energy path has also been plotted superimposed in thick line.}
\end{figure}
%%%%%%%%%%%%%%%%%
The potential energy function $V(R,\theta)$ is taken from the literature, using the {\em ab initio} quantum calculations fitted to an expansion in Legendre polynomials of Essers {\em et al.}~\cite{LiCN.PES}. As can be seen in Fig.~\ref{fig:pes_mep}, the potential energy function has two minima: a relative minimum at $\theta=0$, corresponding to the Li-CN isomer, and an absolute minimum at $\theta=\pi$ rad, corresponding to the most stable CN-Li isomer. Both minima are connected by the minimum energy path (MEP), also represented in Fig.~\ref{fig:pes_mep}. In order to have a suitable analytical expression for some calculations, we have fitted the MEP to an even Fourier series expansion
\begin{equation}
\label{eq:MEP.expansion}
R_\text{eq}(\theta) = \sum_{k=0}^9 C_k\cos(k\theta),
\end{equation}
where the coefficients $C_k$ are given in Table~\ref{tab:MEP.coeff}. Observe that, as is shown in the potential energy profile along the MEP [i.e., $V_\text{eq}(\theta)=V\bm{(}R_\text{eq}(\theta),\theta\bm{)}$] depicted in Fig.~\ref{fig:pes_mep}~(a), the well around the absolute minimum (CN-Li isomer) is very anharmonic. Consequently, the transition to (classical) chaos in this system~\cite{Arranz.LiCN.probability.chaos,Arranz.LiCN.bifurcation.chaos} takes place for energies around 1700 cm$^{-1}$, well below the isomerization barrier energy of 3455 cm$^{-1}$.
%%%%%%%%%%%%%%%%
\begin{table}[t]
\caption{\label{tab:MEP.coeff}%
Numerical values of the fitting coefficients for the minimum energy path series expansion in Eq.~(\ref{eq:MEP.expansion}).}
\begin{ruledtabular}
%\begin{tabular}{crc|cr}
\begin{tabular}{crccr}
$k$ & \multicolumn{1}{c}{$C_k$ (a.u.)} & & $k$ & \multicolumn{1}{c}{$C_k$ (a.u.)} \\
\hline
0 & $ 4.132892896462\phantom{\times10^{-1}}$ & & 5 & $ 6.220709233536$$\times$$10^{-3}$ \\
1 & $ 2.343663054880$$\times$$10^{-1}$       & & 6 & $ 1.514561578552$$\times$$10^{-2}$ \\
2 & $ 4.852236115677$$\times$$10^{-1}$       & & 7 & $-1.780658985867$$\times$$10^{-3}$ \\
3 & $-1.603718738637$$\times$$10^{-2}$       & & 8 & $-4.312497534258$$\times$$10^{-3}$ \\
4 & $-5.879701202854$$\times$$10^{-2}$       & & 9 & $ 2.551825016959$$\times$$10^{-4}$ \\
\end{tabular}
\end{ruledtabular}
\end{table}
%%%%%%%%%%%%%%%%

\subsection{\label{sec:class_calculations}Classical calculations}

The canonical equations of motion corresponding to the Hamiltonian function in Eq.~(\ref{eq:Hamiltonian}) were built, and standard numerical integration used to calculate trajectories for the molecular system. In order to get an illustrative graphical representation of the trajectories in phase space, and especially the resonance structures, we have computed composite Poincar\'e surfaces of section (PSS) across the MEP for different energies. In this representation, classical resonances appear as chains of islands, such that the number of islands indicates the corresponding order of resonance, i.e., the number of oscillations in the associated coordinate along the corresponding periodic orbit.

For this purpose, we have applied the canonical transformation
\begin{align}
\label{eq:PSS_coordinates}
  \rho &= R - R_\text{eq}(\theta), &   \vartheta &= \theta,                                              \nonumber\\
P_\rho &= P_R,                     & P_\vartheta &= P_\theta + P_R \frac{dR_\text{eq}(\theta)}{d\theta},
\end{align}
where $R_\text{eq}(\theta)$ is the series expansion that fits the MEP given in Eq.~(\ref{eq:MEP.expansion}). In this way, for a given energy $E$, the PSS along the MEP is defined in $(\vartheta,P_\vartheta)$ coordinates by taking $\rho=0$ and choosing an arbitrary branch (the negative one in our calculations) in the second degree equation for $P_\rho$ that arises from the Hamiltonian conservation condition $H(\rho,\vartheta,P_\rho,P_\vartheta)=E$.

Additionally, in order to enhance the graphical representation, the center of symmetry existing at $(\vartheta,P_\vartheta)=(\pi,0)$ has been applied to all calculated points, thus doubling the number of points depicted in the corresponding figure.

On the other hand, the calculation of initial conditions, for the main periodic orbits represented in the PSS, was performed following the systematic numerical method described in Ref.~\cite{Arranz.LiCN.bifurcation.chaos}.

\subsection{\label{sec:quant_calculations}Quantum calculations}

Quantum eigenergies and eigenstates of the Hamiltonian operator, corresponding to the Hamiltonian function in Eq.~(\ref{eq:Hamiltonian}), were calculated by means of the {\em discrete variable representation-distributed Gaussian basis} (DVR-DGB) method of Ba\v{c}i\'{c} and Light~\cite{DVR-DGB}. Using a relatively small basis set, this method provides good accuracy for highly excited vibrational states and, as was shown by Ba\v{c}i\'{c} and Light, it works very well for our system.

Considering a single value of the Planck constant, normally $\hbar=1$ a.u., only a few quantum resonances are observed. However, it has been shown in the literature~\cite{Arranz.LiCN.hbar.correlation1,Arranz.LiCN.hbar.correlation2,Arranz.LiCN.HCN.HO2.hbar.correlation,Parraga.KCN.hbar.correlation} that by expanding the range of $\hbar$ values in the calculations, obtaining the correlation diagram of eigenenergies versus Planck constant, others quantum resonances are unveiled. In particular, a singular series of quantum resonances formed by {\em broad} avoided crossings is observed, which constitutes the so-called frontier of scars that separates the regions of order and chaos in quantum systems.

In this way, the DVR-DGB method was applied at values $\hbar=\{0.01,0.02,\ldots,3.00\}\ \text{a.u.}$, obtaining the 130 low lying eigenstates for each value of $\hbar$ with its eigenenergies converged to within 1~cm$^{-1}$. Let us indicate that, in order to maintain accuracy, the number of {\em rays} (the fixed values of $\theta$-coordinate taken in DVR-DGB method) must be increased as $\hbar$ decreases. Thus, a final basis set of 414-418 ray eigenvectors (depending on the value of $\hbar$) lying in 45 rays was used in the range $\hbar\in[1.01,3.00]\ \text{a.u.}$, a basis set of 820-841 ray eigenvectors lying in 90 rays was used in the range $\hbar\in[0.31,1.00]\ \text{a.u.}$, and a basis set of 1480-1710 ray eigenvectors lying in 180 rays was used in the range $\hbar\in[0.01,0.30]\ \text{a.u.}$

\section{\label{sec:results.discussion}Results and discussion}

\subsection{\label{sec:class_resonances}Classical resonances}

According to the theorems of Kolmogorov-Arnold-Moser and Poincar\'e-Birkhoff~\cite{Gutzwiller.caos,Reichl.caos,Lichtenberg.Lieberman.caos}, as energy increases, our system shows a regular behavior up to the order-chaos threshold energy (around 1700 cm$^{-1}$), while different bifurcations take place, giving rise to the corresponding resonances.

\begin{figure}[t]
\includegraphics{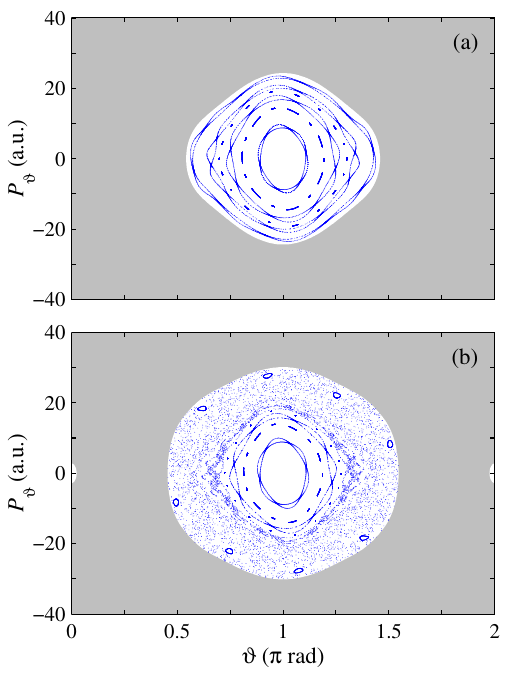}
\caption{\label{fig:pss}%
Composite Poincar\'e surfaces of section, defined along the minimum energy path, for energies of 1500 cm$^{-1}$ (a) and 2300 cm$^{-1}$ (b), below and above, respectively, the order-chaos threshold energy. The chains of islands corresponding to the main resonances have been specifically represented, namely, 1:6, 1:7, 1:8, 1:9, 1:10, again 1:10, and 1:8 (in the chaotic sea) classical resonances. Gray region represents the energetically forbidden region.}
\end{figure}

The main resonances that arise as energy increases correspond to 1:6, 1:7, 1:8, 1:9, 1:10, and another 1:10 resonances. Figure~\ref{fig:pss}~(a) shows the chains of islands associated to each of them in the composite PSS representation. For even resonances ($a$:$b=1$:6, 1:8, 1:10), the corresponding separatrix has been represented in the figure, such that the chain of $b$ islands is clearly visible in each case. However, for odd resonances ($a$:$b=1$:7, 1:9), since the chain of islands is extremely narrow, a set of invariant tori has been represented instead. That is to say, the sets of apparent segments in the figure correspond to sets of extremely narrow invariant tori. Observe that the number of these invariant tori (apparent segments) in each case is $2b$ instead of $b$. This is because there are two stable periodic orbits associated to the odd resonances, such that the corresponding chains of islands appear intercalated. Additional details about the bifurcations arising in this system, including the graphical representation of associated periodic orbits, were reported in Ref.~\cite{Arranz.LiCN.bifurcation.chaos}.
%%%%%%%%%%%%%%%%%
%%%%%%%%%%%%%%%%%

\vspace{2ex}   %to be deleted

In order to quantitatively put in correspondence classical and quantum resonances, we have calculated the energy value of the bifurcation point corresponding to each classical resonance, i.e., the value where, as energy increases, the resonance first take place. These bifurcation energies are listed in Table~\ref{tab:resonance_energies}, where the two 1:10 resonances have been distinguished as 1:10 (i) and 1:10 (ii).
%%%%%%%%%%%%%%%%
\begin{table}[b]
\caption{\label{tab:resonance_energies}%
Energy values (in cm$^{-1}$) for each resonance, corresponding to the classical bifurcation point and the limit $\hbar\to0$ of the series of avoided crossings obtained by different methods: linear extrapolation of the series, semiclassical theory with quadratic fitting (QF), semiclassical theory with coupled quadratic fitting (CQF), semiclassical theory with cubic fitting (CF), and semiclassical theory with coupled cubic fitting (CCF).}
\begin{ruledtabular}
\begin{tabular}{ccccccc}
Resonance & Bifurc. & Extrapol. & QF & CQF & CF & CCF \\
\hline
1:6       &   \phZ205 & \phZ201 & \phZ169 & \phZ165 & \phZ186 & \phZ191 \\
1:7/2:14  &   \phZ588 & \phZ590 & \phZ527 & \phZ599 & \phZ476 & \phZ475 \\
1:8       &   \phZ872 & \phZ860 & \phZ760 & \phZ861 & \phZ704 & \phZ710 \\
1:9/2:18  &      1035 &    1028 & \phZ919 &    1030 & \phZ898 & \phZ926 \\
1:10 (i)  &      1178 &    1141 &    1033 &    1146 &    1083 &    1181 \\
1:10 (ii) &      1327 &    1321 &       - &       - &    1872 &    1544 \\
1:8 (s)   & 1958-2209 &    1840 &       - &       - &    2251 &    1950 \\
\end{tabular}
\end{ruledtabular}
\end{table}
%%%%%%%%%%%%%%%%

Moreover, as energy increases above the order-chaos threshold energy, the onset of chaos takes place, resulting in the breakup of the chains of islands, and leading to the mixed chaos regime in the phase space. In this regime, regular structures can emerge in the chaotic region. Figure~\ref{fig:pss}~(b) shows, in the chaotic sea and close to the energetically forbidden region, a chains of 8 islands corresponding to a 1:8 classical resonance, which is directly related to the frontier of scars in the quantum transition from order to chaos, as has been described in the literature~\cite{Arranz.LiCN.hbar.correlation1,Arranz.LiCN.hbar.correlation2,Arranz.LiCN.HCN.HO2.hbar.correlation}. Notice that the scars are localized over isolated unstable periodic points existing in between the islands. The states localized over the chain of islands, also present in the frontier of scars, are not scars. The generation of the periodic orbits involved in this 1:8 resonance is different and more complex than in the cases described above. In the regular regime, below the order-chaos threshold energy, for each resonance, two periodic orbits (stable and unstable) are generated at a single bifurcation point. However, in the present 1:8 resonance, the involved periodic orbits are generated at two bifurcation points. Consequently, two bifurcation energies are associated to this 1:8 resonance. Additional details about this resonance, including graphical representation of the corresponding periodic orbits, can be obtained from Ref.~\cite{Arranz.LiCN.bifurcation.chaos}.

As in the case of the resonances in the regular regime, we have calculated the two energy values of both bifurcation points related to this particular 1:8 resonance. These bifurcation energies are shown in Table~\ref{tab:resonance_energies}, where the notation 1:8 (s) has been used in reference to the quantum scarring phenomena associated to this resonance.

\subsection{\label{sec:quant_resonances}Quantum resonances}

%%%%%%%%%%%%%%%%%
\begin{figure}[t]
\includegraphics{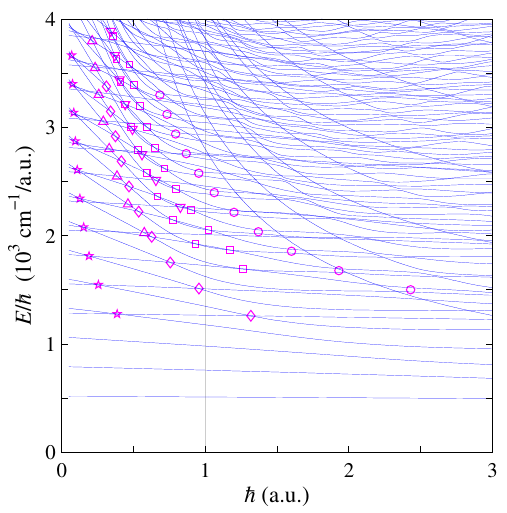}
\caption{\label{fig:correlation_diagram}%
Correlation diagram of eigenenergies versus Planck constant. On grounds of graphical clarity, energy is divided by Planck constant. Different series of avoided crossings, corresponding to 1:6 ($\bigstaropen$), 2:14 ($\bigtriangleup$), 1:8 ($\lozenge$), 2:18 ($\bigtriangledown$), 1:10 ($\square$), and 1:8 ($\bigcirc$) quantum resonances, have been marked.}
\end{figure}
%%%%%%%%%%%%%%%%%
%%%%%%%%%%%%%%%%
\begin{table}[b]
\caption{\label{tab:resonances}%
Quantum numbers $(n_1,n_2)$ and $(n_1^\prime,n_2^\prime)$ corresponding to the states involved in the series of $|n_1 - n_1^\prime|$:$|n_2 - n_2^\prime|$ quantum resonances marked in the correlation diagram represented in Fig.~\ref{fig:correlation_diagram}. The resonances belonging to the same series are identified with the index $k$.}
\begin{ruledtabular}
\begin{tabular}{ccccc}
Resonance & Symbol & $(n_1,n_2)$ & $(n_1^\prime,n_2^\prime)$ & $k$ \\
\hline
1:6  & $\bigstaropen$     & $(0,6+2k)$  & $(1,2k)$ & $0,1,\ldots,9$  \\
2:14 & $\bigtriangleup$   & $(0,14+2k)$ & $(2,2k)$ & $0,1,\ldots,7$  \\
1:8  & $\lozenge$         & $(0,8+2k)$  & $(1,2k)$ & $0,1,\ldots,9$  \\
2:18 & $\bigtriangledown$ & $(0,18+2k)$ & $(2,2k)$ & $1,2,\ldots,8$  \\
1:10 & $\square$          & $(0,10+2k)$ & $(1,2k)$ & $2,3,\ldots,12$ \\
1:8  & $\bigcirc$         & $(0,8+2k)$  & $(1,2k)$ & $2,3,\ldots,12$ \\
\end{tabular}
\end{ruledtabular}
\end{table}
%%%%%%%%%%%%%%%%
%%%%%%%%%%%%%%%%%
\begin{figure*}[t]
\includegraphics{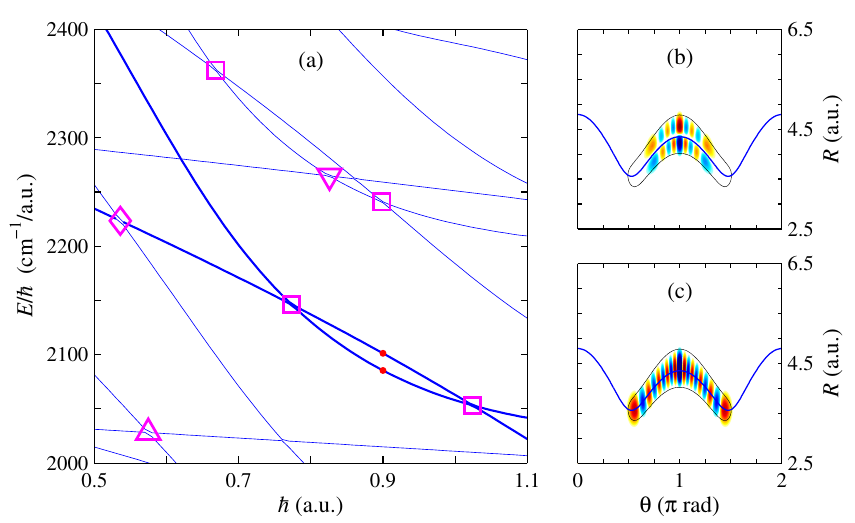}
\caption{\label{fig:detail_r1-10}%
(a) Magnification of the region of the correlation diagram depicted in Fig.~\ref{fig:correlation_diagram} where take place the avoided crossings corresponding to the pair of 1:10 resonances with $k=4$ (see Table~\ref{tab:resonances}). The involved states have been represented in thick line. The corresponding wavefunctions at $\hbar=0.9$ a.u. (value between both avoided crossings) are depicted in panel (b) for the upper state, with quantum numbers $(n_1,n_2)=(1,8)$, and in panel (c) for the lower state, with quantum numbers $(n_1,n_2)=(0,18)$. In both cases, the minimum energy path and the corresponding eigenenergy contour have been plotted superimposed in thick line and thin line, respectively.}
\end{figure*}
%%%%%%%%%%%%%%%%%
The correlation diagram of eigenenergies versus Planck constant calculated for this molecular system, which has been shown to be a very useful tool for the study of quantum transition from order to chaos~\cite{Arranz.LiCN.hbar.correlation1,Arranz.LiCN.hbar.correlation2,Arranz.LiCN.HCN.HO2.hbar.correlation,Parraga.KCN.hbar.correlation}, is depicted in Fig.~\ref{fig:correlation_diagram}. Notice that, in order to obtain a suitable graphical representation for the eigenenergy lines, energy is divided by Planck constant, such that the harmonic behavior is represented by a horizontal line. The frontier of scars defined in Refs.~\cite{Arranz.LiCN.hbar.correlation1,Arranz.LiCN.hbar.correlation2}, which separates the regions of order and chaos, has been marked with open circles. Observe that, according to the results of the random matrix theory~\cite{Gutzwiller.caos,Reichl.caos}, the correlation diagram shows large level repulsion above the frontier of scars (resulting in large and broad avoided crossings), while the repulsion below is extremely small (resulting in narrow and sharp avoided crossings). However, since this system exhibits mixed chaos, level repulsion is not complete, such that regular states exist above the frontier of scars. In particular, we can observe in the mixed chaos region different hyperbolic curves, which correspond to the regular eigenstates located on the relative minimum of the potential energy function at $\theta=0$ (Li-CN isomer). Moreover, observe that, in the regular region (below the frontier of scars), the curves of the eigenenergies follow a pattern, such that we can assign quantum numbers to each eigenstate without having to resort to the nodal structure of the wavefunctions. The excitation in the $\vartheta$-coordinate requires low energy and is very anharmonic. Consequently, as energy increases, the first lines correspond to excitations in the $\vartheta$-coordinate, which are curved and quickly take an increasing negative slope. On the other hand, the excitation in the $\rho$-coordinate requires more energy and is little anharmonic. Thus, the lines corresponding to excitations in the $\rho$-coordinate are widely spaced, not curved (quasi-straight), and slowly take an increasing negative slope (quasi-horizontal). For example, at $\hbar=1$ a.u., the first 14 states correspond to the following quantum numbers: $(n_1,n_2)=(0,0)$, $(0,2)$, $(0,4)$, $(0,6)$, $(1,0)$, $(0,8)$, $(0,10)$, $(1,2)$, $(0,12)$, $(1,4)$, $(0,14)$, $(0,16)$, $(1,6)$, $(2,0)$, where $n_1$ represents excitation in the $\rho$-coordinate and $n_2$ represents excitation in the $\vartheta$-coordinate. Additional details about this correlation diagram and the quantum transition order-chaos can be obtained in Refs.~\cite{Arranz.LiCN.hbar.correlation1,Arranz.LiCN.hbar.correlation2}.

As has been previously shown~\cite{Arranz.LiCN.hbar.correlation1,Arranz.LiCN.hbar.correlation2}, the frontier of scars is formed by a series of {\em broad} avoided crossings corresponding to a 1:8 quantum resonance, such that it is the quantum counterpart of the 1:8 classical resonance shown in Fig.~\ref{fig:pss}~(b), as well as other related 1:8 resonances corresponding to unstable periodic orbits not shown in the figure. Moreover, below the frontier, in the region of order, we have identified different series of {\em sharp} avoided crossings corresponding to 1:10, 2:18, 1:8, 2:14, and 1:6 quantum resonances. All of them are marked with different symbols in Fig.~\ref{fig:correlation_diagram}. The quantum numbers of the pair of eigenstates involved in each resonance are given in Table~\ref{tab:resonances}, where resonances belonging to the same series are identified with the index $k$. Note that, as index $k$ increases, the series goes to $\hbar=0$. As a clarifying example, the detail of the $k=4$ case in the double series of 1:10 resonances has been depicted in Fig.~\ref{fig:detail_r1-10}. Observe that, in the case highlighted in the figure, the eigenstate with 0 and 18 nodal lines, with respect to the MEP [quantum numbers $(n_1,n_2)=(0,18)$], interacts twice with the eigenstate with 1 and 8 nodal lines [quantum numbers $(n_1^\prime,n_2^\prime)=(1,8)$], resulting in both $|n_1-n_1^\prime|$:$|n_2-n_2^\prime|=1$:$10$ resonances with $k=4$.

In view of these results, we can conjecture that the quantum counterpart of each one of the 1:6, 1:7, 1:8, 1:9, 1:10 (i), and 1:10 (ii) classical resonances is the corresponding series of 1:6, 2:14, 1:8, 2:18, and 1:10 (double) quantum resonances identified in the correlation diagram of eigenenergies versus Planck constant. It is worth noting the usefulness of this diagram, since only considering the value $\hbar=1$ a.u. these quantum resonances cannot be observed. Notice that, in this system, some series of resonances begin for $\hbar<1$ a.u.

%%%%%%%%%%%%%%%%%
\begin{figure}[t]
\includegraphics{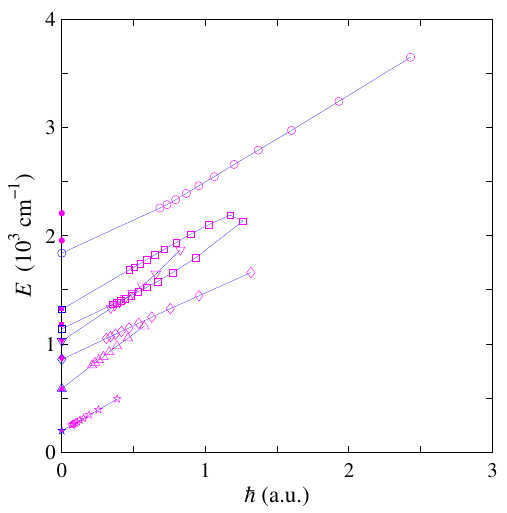}
\caption{\label{fig:resonance_series}%
Series of avoided crossing points, corresponding to 1:6 ($\bigstaropen$), 2:14 ($\bigtriangleup$), 1:8 ($\lozenge$), 2:18 ($\bigtriangledown$), 1:10 ($\square$), and 1:8 ($\bigcirc$) quantum resonances. The points belonging to the same series have been connected with straight lines. The open symbols plotted at $\hbar = 0$ are the result of a linear extrapolation. The filled symbols plotted at $\hbar = 0$ represent the bifurcation energies of the corresponding 1:6 ($\bigstar$), 1:7 ($\blacktriangle$), 1:8 ($\blacklozenge$), 1:9 ($\blacktriangledown$), 1:10 ($\blacksquare$), again 1:10 ($\blacksquare$), 1:8 ($\blackbigcirc$), and again 1:8 ($\blackbigcirc$) classical resonances.}
\end{figure}
%%%%%%%%%%%%%%%%%
Moreover, although the representation $E/\hbar$ versus $\hbar$ for the correlation diagram is suitable in order to obtain the clearest representation of the eigenenergy lines, some features of the series of resonances could be masked. Thus, in order to reveal these features, the series of avoided crossing points corresponding to all resonances have been depicted in Fig.~\ref{fig:resonance_series} using the $E$ versus $\hbar$ representation. First, we can observe that two series lines may cross, as is the case of 2:18 and 1:10 resonances series. Consequently, the quantum-classical correspondence in the order of appearance of the resonances as energy increases will only be held in a range of decreasing $\hbar$ values, or more generally, in the limit $\hbar\to0$. Also, we can observe in Fig.~\ref{fig:resonance_series} that the behavior of the series lines is mostly linear, i.e., the avoided crossing points lie approximately along straight lines, albeit some curvature is visible in certain cases. This mostly linear behavior allow us to calculate numerically the semiclassical limit $\hbar\to0$ of the resonances series by means of linear extrapolation. The corresponding energy values obtained are given in Table~\ref{tab:resonance_energies} and also represented, along with the bifurcation energies of the classical resonances, in Fig.~\ref{fig:resonance_series}. Notice that a clear correspondence between classical bifurcation energies and quantum series energies in the semiclassical limit can be established for resonances 1:6, 2:14, 1:8, 2:18, and 1:10 (double). This is one of the main results of this paper. The small differences between classical and quantum (semiclassical) energies are due to the deviation from the linear behavior of the resonance lines, singularly in the 1:10 (i) resonance, leading to small errors in the numerical calculations through linear extrapolation.

On the other hand, the 1:8 (s) resonance case is somewhat different. As was described above, two bifurcation energies are associated to this resonance, consequently we have an energy interval, rather than an energy point, to put in correspondence with the quantum energy. That is, we would expect that the semiclassical limit of the 1:8 (s) resonances series belongs to this energy interval. However, the energy value obtained through linear extrapolation is outside (below) the energy interval. As in the 1:10 (i) resonance case, this disagreement may be caused by the deviation from the linear behavior of the resonance line. Indeed, as can be observed in Fig.~\ref{fig:resonance_series}, the $k=12$ last resonance represented in the 1:8 (s) resonances series clearly deviates from the linear behavior of the other resonances in the series. Therefore, we could assume that this deviation continues as $k$ increases, such that, in the semiclassical limit, the energy value belongs to the interval. In any case, due to its singular characteristics, the 1:8 (s) resonances series requires further research.

\subsection{\label{sec:semiclassical}Semiclassical theory}

%%%%%%%%%%%%%%%%%
\begin{table*}[t]
\caption{\label{tab:fitting_coefficients}%
Values of the coefficients corresponding to the series expansion used in the semiclassical theory for different fitting levels, namely, quadratic fitting (QF), coupled quadratic fitting (CQF), cubic fitting (CF), and coupled cubic fitting (CCF).}
\begin{ruledtabular}
\begin{tabular}{cccccccccc}
Fitting & $\alpha_1$ & $\alpha_2$ & $\beta_{11}$\footnotemark[1] & $\beta_{12}$ & $\beta_{22}$\footnotemark[1] & $\gamma_{111}$\footnotemark[1] & $\gamma_{112}$ & $\gamma_{122}$ & $\gamma_{222}$\footnotemark[1] \\
level   & (cm$^{-1}\!$/a.u.) & (cm$^{-1}\!$/a.u.) & (cm$^{-1}\!$/a.u.$\!^2$) & (cm$^{-1}\!$/a.u.$\!^2$) & (cm$^{-1}\!$/a.u.$\!^2$) & (cm$^{-1}\!$/a.u.$\!^3$) & (cm$^{-1}\!$/a.u.$\!^3$) & (cm$^{-1}\!$/a.u.$\!^3$) & (cm$^{-1}\!$/a.u.$\!^3$) \\
\hline
QF  & 786.226 & 139.005 &           15.039 &            - & 3.180 &            - &            - &            - &     - \\
CQF & 796.124 & 138.848 &        \phZ4.871 &        6.257 & 3.048 &            - &            - &            - &     - \\
CF  & 780.137 & 142.085 &        \phZ6.617 &            - & 4.608 & $-$5.601\phM &            - &            - & 0.097 \\
CCF & 774.431 & 142.068 & \phZ$-$2.310\phM & $-$3.144\phM & 4.621 &        3.634 & $-$3.677\phM & $-$0.311\phM & 0.104 \\
\end{tabular}
\end{ruledtabular}
\footnotetext[1]{Note that, for non-coupled series expansions (i.e., QF and CF), coefficient index notation has been simplified in the equations in text, such that $\beta_{11}\equiv\beta_1$, $\beta_{22}\equiv\beta_2$, $\gamma_{111}\equiv\gamma_1$, and $\gamma_{222}\equiv\gamma_2$.}
\end{table*}
%%%%%%%%%%%%%%%%%
In order to develop a semiclassical theory for the energy corresponding to the quantum resonances series in the limit $\hbar\to0$, we will consider the $\hbar$ series expansion
\begin{equation}
\label{eq:E_hbar}
E(\hbar) = c_1\hbar - c_2\hbar^2 + c_3\hbar^3 + \cdots,
\end{equation}
with $c_1,c_2,c_3>0$, as an approximation to the energy levels in the regular region of the correlation diagram. Observe that, as is clear from Fig.~\ref{fig:correlation_diagram}, where $E/\hbar$ vs $\hbar$ rather than $E$ vs $\hbar$ is represented, first and third order coefficients should be positive, while second order coefficient should be negative. Accordingly, we have explicitly written the corresponding signs in Eq.~(\ref{eq:E_hbar}). Moreover, quantum numbers $(n_1,n_2)$ will be included in the coefficients $c_m$ with the form $(n_i + 1/2)^m$, assuming the topological index of 2 for both degrees of freedom, following the customary expansion of Fues~\cite{Fues,Morse}. Also, coupled factors $(n_1 + 1/2)^{m_1}(n_2 + 1/2)^{m_2}$, with $m_1+m_2=m$, will be considered. Notice that coupled expansions are the strictly correct expressions for the $\hbar$ series expansions, and the corresponding decoupled expansions should only be taken as a conjectured approximation in order to obtain simpler expressions.

Thus, applying Eq.~(\ref{eq:E_hbar}), the crossing between two states with quantum numbers $(n_1,n_2)$ and $(n_1^\prime,n_2^\prime)$ will be given by the condition $E(\hbar_0)=E^\prime(\hbar_0)$. According to the quantum resonance data in Table~\ref{tab:resonances}, the quantum numbers involved in the $a$:$b$ resonances series are determined by the index $k$, such that $(n_1,n_2)=(0,b+2k)$ and $(n_1^\prime,n_2^\prime)=(a,2k)$. Hence, the crossing condition will be also determined by the index $k$, that is, $E(\hbar_k)=E^\prime(\hbar_k)$, leading to the series of crossing points $(\hbar_k,E_k)$ for the $a$:$b$ resonance. We will use the weighted least squares method, applied on these series of crossing points and the series of avoided crossing points obtained from the correlation diagram, to procure the corresponding fitting coefficients that determine the series expansion. In order to enhance the relevant region for the semiclassical limit $\hbar\to0$, the fitting weight $1/\hbar^3$ is used.

Eventually, the energy of the $a$:$b$ resonances series in the limit $\hbar\to0$ is obtained by calculating the limit $\lim_{k\to\infty}E_k=E_\infty$ for the corresponding crossing points in the series expansion. In all cases, the corresponding $\hbar$ limit is zero, i.e., $\lim_{k\to\infty}\hbar_k=0$, hence, for the resonances series, the limit $k\to\infty$ implies the limit $\hbar\to0$.

Next, the results obtained for different (coupled and decoupled) expansion orders are shown.

\subsubsection{\label{sec:QF}Quadratic series expansion}

For the quadratic series expansion
\begin{align}
\label{eq:QF_expansion}
E^\text{QF}(\hbar) &= \left[ \alpha_1\left( n_1 + \frac{1}{2} \right) + \alpha_2\left( n_2 + \frac{1}{2} \right) \right] \hbar \nonumber\\
                   &\quad - \left[ \beta_1\left( n_1 + \frac{1}{2} \right)^2 + \beta_2\left( n_2 + \frac{1}{2} \right)^2 \right] \hbar^2,
\end{align}
where $\alpha_i$ and $\beta_i$ ($i=1,2$) are the coefficients obtained from the corresponding quadratic fitting (QF), which are given in Table~\ref{tab:fitting_coefficients}, the energy of the $a$:$b$ resonances series in the limit $\hbar\to0$ is
\begin{equation}
\label{eq:QF_limit}
E_\infty^\text{QF} = \frac{(b\alpha_2)^2 - (a\alpha_1)^2}{(2b)^2\beta_2}.
\end{equation}
The energies given by Eq.~(\ref{eq:QF_limit}) for the different resonances series studied in this paper are listed in Table~\ref{tab:resonance_energies}. Assuming the classical bifurcation energy as the correct value, the mean absolute relative difference for all resonances is 0.127, with a standard deviation of 0.027.

Observe that no results appear in Table~\ref{tab:resonance_energies} for resonances 1:10 (ii) and 1:8 (s). Indeed, the QF approximation can only account for resonances corresponding to first order crossings, i.e., the first crossing between two states as $\hbar$ increases. Note that, in $E/\hbar$ vs $\hbar$ representation of the correlation diagram, QF approximation results in straight lines, and consequently they cross only once. As we can observe in the correlation diagram depicted in Fig.~\ref{fig:correlation_diagram}, the 1:6, 2:14, 1:8, 2:18, and 1:10 (i) resonances series correspond approximately to avoided crossings of straight lines. However, this is not the case for 1:10 (ii) and 1:8 (s) resonances, which correspond to second order crossings, i.e., the second crossing between two states as $\hbar$ increases. As is clear from Fig.~\ref{fig:detail_r1-10}~(a), where the crossings corresponding to 1:10 (i) and 1:10 (ii) resonances (with $k=4$) are depicted in detail, the existence of a second crossing requires a curved line. This will be allowed by the cubic series expansion.

Finally, it is worth noting that the quadratic series expansion is equivalent to the well-known Morse oscillator~\cite{Morse} for two degrees of freedom, namely
\begin{equation}
\label{eq:Morse_levels}
E^\text{Mo}(\hbar) = \hbar\Omega_1(\hbar) \left( n_1 + \frac{1}{2} \right) + \hbar\Omega_2(\hbar) \left( n_2 + \frac{1}{2} \right),
\end{equation}
where
\begin{equation}
\label{eq:Morse_frequency}
\Omega_i(\hbar) = \omega_i \left[ 1 - \frac{1}{4D_i}\hbar\omega_i \left( n_i + \frac{1}{2} \right) \right]
\end{equation}
(with $i=1,2$) are the state dependent Morse frequencies, and $\omega_i,D_i>0$ are the harmonic frequency and energy well depth for the $i$-th degree of freedom, respectively. Using the Morse notation, the energy of the $a$:$b$ resonances series in the limit $\hbar\to0$ given in Eq.~(\ref{eq:QF_limit}) becomes
\begin{equation}
\label{eq:Morse_limit}
E_\infty^\text{Mo} = D_2 \left[ 1 - \left( \frac{a}{b}\ \frac{\omega_1}{\omega_2} \right)^2 \right].
\end{equation}
This expression is interesting since it makes physical sense. First, it depends only on the energy well depth $D_2$. This feature reflects the fact that the series of avoided crossings are mostly due to the anharmonic behavior of the excitations in the coordinate perpendicular to MEP (approximately, the $\theta$-coordinate). Moreover, the expression depends on the ratio $(a\text{:}b)/(\omega_2\text{:}\omega_1)$. Since the relation $a\text{:}b\equiv|n_1 - n_1^\prime|$:$|n_2 - n_2^\prime|$ corresponds exactly with the frequency relation $\omega_2^\text{HO}\text{:}\omega_1^\text{HO}$ for the harmonic oscillator~%
%%%%%%%%%
\footnote{Given the energy expression for the harmonic oscillator $E(\hbar)=[\omega_1^\text{HO}(n_1 + 1/2) + \omega_2^\text{HO}(n_2 + 1/2)]\hbar$, at the crossing $E=E^\prime$, it holds $(n_1 - n_1^\prime)/(n_2^\prime - n_2)=\omega_2^\text{HO}/\omega_1^\text{HO}$.}%
%%%%%%%%%
, the ratio $(a\text{:}b)/(\omega_2\text{:}\omega_1)$ reflects the deviation from harmonic behavior. Note that, in the harmonic limit, where $\omega_2\text{:}\omega_1$ tends to $\omega_2^\text{HO}\text{:}\omega_1^\text{HO}$, the energy $E_\infty^\text{Mo}$ goes to zero. Indeed, for the harmonic oscillator, all states cross at a single point: the origin $(\hbar,E)=(0,0)$.

\subsubsection{\label{sec:CQF}Coupled quadratic series expansion}

The results obtained with the QF level can be enhanced by allowing coupling between both degrees of freedom. In this way, we will consider the coupled quadratic series expansion
\begin{widetext}
\begin{equation}
\label{eq:CQF_expansion}
E^\text{CQF}(\hbar) = \left[ \alpha_1\left( n_1 + \frac{1}{2} \right) + \alpha_2\left( n_2 + \frac{1}{2} \right) \right] \hbar 
                      - \left[ \beta_{11}\left( n_1 + \frac{1}{2} \right)^2 +
                               \beta_{12}\left( n_1 + \frac{1}{2} \right)\left( n_2 + \frac{1}{2} \right) +
                               \beta_{22}\left( n_2 + \frac{1}{2} \right)^2 \right] \hbar^2,
\end{equation}
\end{widetext}
where $\alpha_i$ and $\beta_{ij}$ ($i,j=1,2$) are the coefficients obtained from the corresponding coupled quadratic fitting (CQF) given in Table~\ref{tab:fitting_coefficients}. By using the series expansion in Eq.~(\ref{eq:CQF_expansion}), the energy for the $a$:$b$ resonances series in the limit $\hbar\to0$ will be

\vspace{1ex}   %to be deleted

\begin{equation}
\label{eq:CQF_limit}
E_\infty^\text{CQF} = \frac{(b\alpha_2 - a\alpha_1)(b\alpha_2\beta_{22} + a\alpha_1\beta_{22} - a\alpha_2\beta_{12})}{(2b\beta_{22} - a\beta_{12})^2}.
\end{equation}

\vspace{1ex}   %to be deleted

Note that this CQF expression reduces to the QF expression in Eq.~(\ref{eq:QF_limit}) by taking $\beta_{12}=0$ (and assuming $\beta_{22}\equiv\beta_2$). The energies given by Eq.~(\ref{eq:CQF_limit}) for resonances series corresponding to first order crossings are given in Table~\ref{tab:resonance_energies}. Observe that, as a whole, CQF energies approach the bifurcation values better than QF energies. Thus, the corresponding mean absolute relative difference for all resonances is 0.051, with a standard deviation of 0.079.

\subsubsection{\label{sec:CF}Cubic series expansion}

In order to obtain an analytical expression available for the resonances corresponding to second order crossings, we may consider the cubic series expansion
\begin{align}
\label{eq:CF_expansion}
E^\text{CF}(\hbar) &=       \left[ \alpha_1\left( n_1 + \frac{1}{2} \right) + \alpha_2\left( n_2 + \frac{1}{2} \right) \right] \hbar \nonumber\\
                   &\quad - \left[ \beta_1\left( n_1 + \frac{1}{2} \right)^2 + \beta_2\left( n_2 + \frac{1}{2} \right)^2 \right] \hbar^2, \nonumber\\
                   &\quad + \left[ \gamma_1\left( n_1 + \frac{1}{2} \right)^3 + \gamma_2\left( n_2 + \frac{1}{2} \right)^3 \right] \hbar^3,
\end{align}
where $\alpha_i$, $\beta_i$, and $\gamma_i$ ($i=1,2$) are the coefficients given by the corresponding cubic fitting (CF), which are listed in Table~\ref{tab:fitting_coefficients}. The series expansion in Eq.~(\ref{eq:CF_expansion}) leads to the following expression for the energy of the $a$:$b$ resonances series in the limit $\hbar\to0$
\begin{equation}
\label{eq:CF_limit}
E_\infty^\text{CF} = \frac{\beta_2(b\alpha_2 - a\alpha_1)}{9b\gamma_2}
                     + \rho \left( \alpha_2 - \frac{(b\alpha_2 - a\alpha_1)}{3b} - \frac{2\beta_2^2}{9\gamma_2} \right),
\end{equation}
where
\begin{equation}
\label{eq:CF_limit_root}
\rho = \frac{b\beta_2 \pm \left[ b^2\beta_2^2 - 3b\gamma_2(b\alpha_2 - a\alpha_1) \right]^{1/2} }{3b\gamma_2}.
\end{equation}
Notice that, in this case, two energy values are yielded for the $a$:$b$ resonance, corresponding to take minus sign or plus sign for the root in Eq.~(\ref{eq:CF_limit_root}). For resonances without second order crossings (i.e., $1$:$6$, $2$:$14$, and $2$:$18$), only minus sign makes physical sense. However, for resonances with second order crossings (i.e., $1$:$8$ and $1$:$10$), minus sign gives the energy corresponding to the first crossing, while plus sign gives the energy corresponding to the second crossing. The energies given by Eq.~(\ref{eq:CF_limit}) for all resonances series are given in Table~\ref{tab:resonance_energies}. By considering resonances corresponding to first order crossings, hence comparable with QF and CQF results, the mean absolute relative difference is 0.136, with a standard deviation of 0.053. Moreover, by considering all resonances, the mean absolute relative difference is 0.158, with a standard deviation of 0.118.

\subsubsection{\label{sec:CCF}Coupled cubic series expansion}

As in the case of the QF level results, the results obtained with the CF level can be enhanced by allowing the coupling between both degrees of freedom. Thus, we will consider the coupled cubic series expansion
\begin{widetext}
\begin{align}
\label{eq:CCF_expansion}
E^\text{CCF}(\hbar) &=       \left[ \alpha_1\left( n_1 + \frac{1}{2} \right) + \alpha_2\left( n_2 + \frac{1}{2} \right) \right] \hbar
                           - \left[ \beta_{11}\left( n_1 + \frac{1}{2} \right)^2 +
                                    \beta_{12}\left( n_1 + \frac{1}{2} \right)\left( n_2 + \frac{1}{2} \right) +
                                    \beta_{22}\left( n_2 + \frac{1}{2} \right)^2 \right] \hbar^2 \nonumber\\
                    &\quad + \left[ \gamma_{111}\left( n_1 + \frac{1}{2} \right)^3 +
                                    \gamma_{112}\left( n_1 + \frac{1}{2} \right)^2\left( n_2 + \frac{1}{2} \right) +
                                    \gamma_{122}\left( n_1 + \frac{1}{2} \right)\left( n_2 + \frac{1}{2} \right)^2 +
                                    \gamma_{222}\left( n_2 + \frac{1}{2} \right)^3 \right] \hbar^3,
\end{align}
\end{widetext}
where $\alpha_i$, $\beta_{ij}$, and $\gamma_{ijk}$ ($i,j,k=1,2$) are the coefficients obtained from the corresponding coupled cubic fitting (CCF) given in Table~\ref{tab:fitting_coefficients}. By using the series expansion in Eq.~(\ref{eq:CCF_expansion}), the energy for the $a$:$b$ resonances series in the limit $\hbar\to0$ will be
\begin{align}
\label{eq:CCF_limit}
E_\infty^\text{CCF} &= \left( \frac{p_1}{p_3}\beta_{22} - \frac{p_1p_2}{p_3^2}\gamma_{222} \right) \nonumber\\
                    &\quad - \rho \left[ \alpha_2 - \frac{p_2}{p_3}\beta_{22} + \left( \frac{p_2^2}{p_3^2} - \frac{p_1}{p_3} \right) \gamma_{222} \right],
\end{align}
where
\begin{subequations}
\label{eq:CCF_limit_binoms}
\begin{eqnarray}
\label{eq:CCF_limit_binom_1}
p_1 &=& (a\alpha_1 - b\alpha_2), \\
\label{eq:CCF_limit_binom_2}
p_2 &=& (a\beta_{12} - 2b\beta_{22}), \\
\label{eq:CCF_limit_binom_3}
p_3 &=& (a\gamma_{122} - 3b\gamma_{222}),
\end{eqnarray}
\end{subequations}
and
\begin{equation}
\label{eq:CCF_limit_root}
\rho = \frac{-p_2 \pm (p_2^2 - 4p_1p_3)^{1/2}}{2p_3}.
\end{equation}
Note that the CCF expression reduces to the CF expression in Eq.~(\ref{eq:CF_limit}) by taking $\beta_{12}=\gamma_{122}=0$ (and assuming $\beta_{22}\equiv\beta_2$ and $\gamma_{222}\equiv\gamma_2$). Also, two energy values are yielded for the $a$:$b$ resonance, depending on the sign choice for the root in Eq.~(\ref{eq:CCF_limit_root}). For resonances without second order crossings, only minus sign makes physical sense, while for resonances with second order crossings, minus sign gives the energy corresponding to the first crossing and plus sign gives the energy corresponding to the second crossing. The energies given by Eq.~(\ref{eq:CCF_limit}) for all resonances series are given in Table~\ref{tab:resonance_energies}. Observe that indeed CCF energies approach the bifurcation values better than CF energies. By considering resonances corresponding to first order crossings (comparable with QF and CQF results), the mean absolute relative difference is 0.109, with a standard deviation of 0.080. Besides, by considering all resonances, the mean absolute relative difference is 0.104, with a standard deviation of 0.075.

\section{\label{sec:conclusions}Summary and conclusions}

\vspace{2ex}   %to be deleted

We have studied the correspondence between classical and quantum resonances in the two degrees of freedom isomerizing system CN-Li$\leftrightarrows$Li-CN. For this purpose, the main classical resonances (periodic orbits) have been obtained, being represented as chains of islands in a suitable composite Poincar\'e surface of section, and the corresponding bifurcation energies have been calculated. Moreover, the correlation diagram of eigenenergies versus Planck constant has been obtained, where the main quantum resonances (avoided crossings) have been identified. It is worth noting the usefulness of this diagram, since only considering the value $\hbar=1$ a.u. these quantum resonances would not be observed. Indeed, the physico-chemical relevance of results like those presented here have been discussed in Refs.~\cite{Arranz.LiCN.frontier.spectra1,Arranz.LiCN.frontier.spectra2}.

\vspace{3ex}   %to be deleted

We have established a qualitative correspondence between the classical resonances (as energy increases, 1:6, 1:7, 1:8, 1:9, 1:10, and again 1:10) and the {\em series} of quantum resonances existent in the correlation diagram [1:6, 2:14, 1:8, 2:18, and 1:10 (double)]. These series can cross, such that the quantum-classical correspondence in the order of appearance of the resonances as energy increases is held in a range of decreasing $\hbar$ values, or more generally, in the semiclassical limit $\hbar\to0$. Additionally, for each series, the energy value in the limit $\hbar\to0$ has been numerically calculated, such that we have established a quantitative correspondence between the classical bifurcation energies and the energy of the quantum resonances series in the semiclassical limit.

\vspace{3ex}   %to be deleted

On the other hand, the resonances series corresponding to the so-called frontier of scars (quantum frontier between order and chaos, previously established in the literature) has been also included in our study, leading to partial results. While, due to its singular characteristics (e.g., not one but two bifurcation energies are involved), this resonances series requires further research.

Last, in order to obtain analytical expressions for the energy value of the resonances series in the limit $\hbar\to0$, we have developed a semiclassical theory, considering an $\hbar$ series expansion with different expansion orders (quadratic and cubic), and with coupling and decoupling between both degrees of freedom.

\begin{acknowledgments}
This research was supported by the Ministry of Science, Innovation and Universities-Spain under Grant No.\ PGC2018-093854-B-I00, by ICMAT Severo Ochoa under Grant No.\ CEX2019-000904-S, and by the People Programme (Marie Curie Actions) of the European Union's Horizon 2020 Research and Innovation Program under Grant No.\ 734557.
\end{acknowledgments}

\end{document}